\documentclass{article}
\usepackage[utf8]{inputenc}

\usepackage{hyperref}
\usepackage{amsmath,amsfonts,amssymb}
\usepackage{amsthm}
\usepackage{color}
\usepackage{graphicx}
\usepackage{tikz-cd}
\usetikzlibrary {decorations.pathmorphing,shadows,fadings}

\def\ds{\displaystyle}
\def\bea{\begin{array}{c}}
\def\ea{\end{array}}
\def\be{\begin{equation}\bea\ds}
\def\ee{\ea\end{equation}}
\def\bee{\begin{equation}\begin{array}{rcl}\ds}
\def\eee{\end{array}\end{equation}}

\title{Toy Gravizap for Black Hole Redemption}
\author{Dmitry Melnikov}
\date{}

\begin{document}

\maketitle

\begin{center}
\textit{\small International Institute of Physics, Federal University of 
Rio Grande do Norte, \\ Campus Universit\'ario, Lagoa Nova, Natal-RN  
59078-970, Brazil}

\vspace{2cm}

\end{center}

\vspace{-2cm}

\begin{abstract}
    In this note we discuss a topological realization of the quantum teleportation protocol. Such a realization facilitates keeping track of the evolution of quantum correlations in the system. It is then suggested, that an implementation of a similar protocol can serve as a toy model of a process of information transfer between causally disconnected regions, such as the one that should take place in a unitary evolution of a quantum black hole. Hawking quanta provide the necessary entangled pairs for the teleportation channel, and the same interaction mechanism that creates the Hawking pairs, destroys the entanglement at the late stages of the evolution. Different steps of the teleportation protocol are characterized by the entropies associated with the disconnected regions, and one can introduce an analog of the Page curve for the entropy of the Hawking radiation. Some regions of the ``black hole'' interior appear to actually encode the information of the exterior, so it is natural to identify them with the ``islands'' of the semiclassical gravity analysis.
\end{abstract}

\bigskip
Quantum correlations make possible effects incompatible with our naive notion of locality. The \emph{spooky action at distance} is one example of such an effect~\cite{EPR}. A similar effect is quantum teleportation that allows an instant recreation of a distant quantum state~\cite{NielsenChuang}, although with appropriate preparation and following a classical conference session. With the help of quantum teleportation information can be extracted from a causally disconnected region of space, such as the other side of a two-sided black hole~\cite{Maldacena:2017axo,Gao:2019nyj}.

In this note we will consider a simple model of quantum teleportation in a topological theory. Topological property means, in principle, that the model has no continuous notion of distance in space and consequently no causality. On one hand, this makes quantum teleportation somewhat trivial. However, one should rather consider topological theories as essentially discrete models of space and time -- distance and causality can be recovered in an appropriate limit. So, the model to be considered is rather a model of bits on a lattice, which might have, nevertheless, a continuum formulation as a topological quantum field theory (TQFT), such as Chern-Simons theory. 

We will use the teleportation protocol in the TQFT-grounded bit model to discuss the problem of black hole evaporation and information retrieval from the interior. Recent progress in the solution of this problem~\cite{Penington:2019npb,Almheiri:2019psf,Penington:2019kki,Almheiri:2019qdq}, also known as the black hole information paradox~\cite{Hawking:1976ra}, demonstrated that information can be preserved even at the level of semiclassical gravity analysis, if a corrected prescription for entropy calculation is used. The goal of the present discussion is to reproduce some specific features of the solution in a simple microscopic model, such as bit models of black holes~\cite{Mathur:2009hf,Giddings:2011ks,Giddings:2012dh,Maxfield:2022sio}. In this case the quantum teleportation allows for a specific realization of the Hayden-Preskill protocol~\cite{Hayden:2007cs} of the information retrieval from a black hole. 

The embedding of the discussion in a TQFT framework makes it different from other models studying the same problem. The advantage of this approach is in a simple and intuitive form of the analysis and results. However, the main motivation is the connection between topological theories and low-dimensional models of gravity, such as the example of three-dimensional gravity in anti de Sitter space, which can be cast as a $SO(2,2)$ Chern-Simons theory~\cite{Witten:1989sx}. One may ask whether the prescription of the semiclassical gravity is valid beyond gravity and can be understood in topological theories, in which quantum calculations are under greater control. If yes, then it may be possible to make an explicit connection between the effective semiclassical approach and a full quantum calculation. The simple model considered in this note will produce a unitary behavior of the entropy of Hawking radiation. It will also make rather obvious, how the entropy calculation is modified at late times, due to appearance of regions in the interior that encode the information in the exterior. Analogous regions, called ``islands'', appear in the modified prescription of the semiclassical gravity solution of the information paradox. 

Consider a pair of TQFT states given by the following diagrams:\footnote{To be concrete let these be states of $SU(2)$ Chern-Simons theory in $S^3$ with $S^2$ boundaries~\cite{Witten:1988hf}. See~\cite{WittenTQFT,AtiyahTQFT} for general aspects of TQFT and~\cite{Melnikov:2018zfn} for the specific setup.}
\be
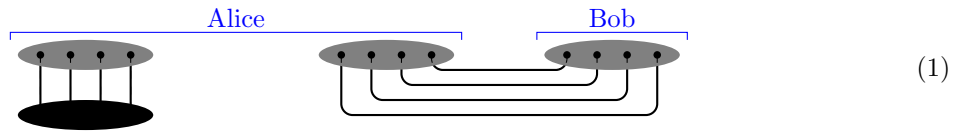
\label{pairofstates}
\begin{tikzpicture}
\draw[thick] (-0.6,0) -- (-0.6,-0.8);
\draw[thick] (-0.2,0) -- (-0.2,-0.8);
\draw[thick] (0.2,0) -- (0.2,-0.8);
\draw[thick] (0.6,0) -- (0.6,-0.8);
\fill[gray] (0,0) ellipse (0.9cm and 0.2cm);
\fill[black] (-0.6,0) circle (0.05);
\fill[black] (-0.2,0) circle (0.05);
\fill[black] (0.2,0) circle (0.05);
\fill[black] (0.6,0) circle (0.05);
\draw[dashed] (-0.6,0) -- (-0.6,-0.8);
\draw[dashed] (-0.2,0) -- (-0.2,-0.8);
\draw[dashed] (0.2,0) -- (0.2,-0.8);
\draw[dashed] (0.6,0) -- (0.6,-0.8);
\fill[black] (0,-0.8) ellipse (0.9cm and 0.2cm);
\newcommand{\x}{4}
\draw[thick] (-0.6+\x,0) -- (-0.6+\x,-0.65) arc (180:270:0.15) -- (0.45+\x+3,-0.8) arc (-90:0:0.15) -- (0.6+\x+3,0);
\draw[thick] (-0.2+\x,0) -- (-0.2+\x,-0.45) arc (180:270:0.15) -- (0.05+\x+3,-0.6) arc (-90:0:0.15) -- (0.2+\x+3,0);
\draw[thick] (0.2+\x,0) -- (0.2+\x,-0.25) arc (180:270:0.15) -- (-0.35+\x+3,-0.4) arc (-90:0:0.15) -- (-0.2+\x+3,0);
\draw[thick] (0.6+\x,0) -- (0.6+\x,-0.05) arc (180:270:0.15) -- (-0.75+\x+3,-0.2) arc (-90:0:0.15) -- (-0.6+\x+3,0);
\fill[gray] (0+\x,0) ellipse (0.9cm and 0.2cm);
\fill[black] (-0.6+\x,0) circle (0.05);
\fill[black] (-0.2+\x,0) circle (0.05);
\fill[black] (0.2+\x,0) circle (0.05);
\fill[black] (0.6+\x,0) circle (0.05);
\draw[dashed] (-0.6+\x,0) -- (-0.6+\x,-0.2);
\draw[dashed] (-0.2+\x,0) -- (-0.2+\x,-0.2);
\draw[dashed] (0.2+\x,0) -- (0.2+\x,-0.2);
\draw[dashed] (0.6+\x,0) -- (0.6+\x,-0.05) arc (180:270:0.15);
\draw[blue] (-1,0.2) -- (-1,0.3) -- (\x+1,0.3) -- (\x+1,0.2);
\draw[blue] (\x/2,0.5) node {Alice};
\renewcommand{\x}{7}
\fill[gray] (0+\x,0) ellipse (0.9cm and 0.2cm);
\fill[black] (-0.6+\x,0) circle (0.05);
\fill[black] (-0.2+\x,0) circle (0.05);
\fill[black] (0.2+\x,0) circle (0.05);
\fill[black] (0.6+\x,0) circle (0.05);
\draw[dashed] (-0.6+\x,0) -- (-0.6+\x,-0.05) arc (0:-90:0.15);
\draw[dashed] (-0.2+\x,0) -- (-0.2+\x,-0.2);
\draw[dashed] (0.2+\x,0) -- (0.2+\x,-0.2);
\draw[dashed] (0.6+\x,0) -- (0.6+\x,-0.2);
\draw[blue] (\x-1,0.2) -- (\x-1,0.3) -- (\x+1,0.3) -- (\x+1,0.2);
\draw[blue] (\x,0.5) node {Bob};
\end{tikzpicture}
\ee
These diagrams should be understood as follows: The gray disks are two-dimensional boundaries of three-dimensional topological spaces; These three-dimensional spaces contain codimension-two defects (Wilson lines) that extend between the same or different boundaries -- their precise topological configuration and other topological details of the three-dimensional space define which precise quantum state one is dealing with; The black blob denotes an unspecified tangle of the lines. The diagrams completely define the states. Alternatively the latter can be defined as functional integrals of the respective TQFT in the respective three-dimensional space with the respective two-dimensional boundary and a corresponding insertion of Wilson line operators. A review of the simple diagrammatic calculus can be found in~\cite{Melnikov:2022qyt}.   

In~(\ref{pairofstates}) one can think of the first state as an arbitrary state of a qubit, while the second as a pair of maximally entangled qubits~\cite{Melnikov:2018zfn}. The first state and one of the entangled qubits is in Alice's possession, while Bob has the second entangled qubit. In order for Alice to teleport the unspecified qubit to Bob she needs to apply an entangling operation on her two qubits, measure both of them and communicate the results to Bob. Depending on those results Bob can recover the original Alice's state through a unitary transformation on his qubit~\cite{Bennett:1992tv,NielsenChuang}.

In the TQFT setting, the simplest way for Alice to achieve the result would be to apply an operation shown in the following diagram:
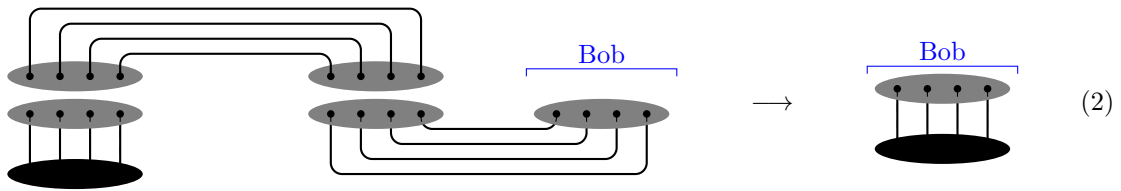
\be
\label{projection1}
\begin{array}{c}
\begin{tikzpicture}
\draw[thick] (-0.6,0) -- (-0.6,-0.8);
\draw[thick] (-0.2,0) -- (-0.2,-0.8);
\draw[thick] (0.2,0) -- (0.2,-0.8);
\draw[thick] (0.6,0) -- (0.6,-0.8);
\fill[gray] (0,0) ellipse (0.9cm and 0.2cm);
\fill[black] (-0.6,0) circle (0.05);
\fill[black] (-0.2,0) circle (0.05);
\fill[black] (0.2,0) circle (0.05);
\fill[black] (0.6,0) circle (0.05);
\draw[dashed] (-0.6,0) -- (-0.6,-0.8);
\draw[dashed] (-0.2,0) -- (-0.2,-0.8);
\draw[dashed] (0.2,0) -- (0.2,-0.8);
\draw[dashed] (0.6,0) -- (0.6,-0.8);
\fill[black] (0,-0.8) ellipse (0.9cm and 0.2cm);
\newcommand{\y}{0.5}
\fill[gray] (0,0+\y) ellipse (0.9cm and 0.2cm);
\fill[black] (-0.6,0+\y) circle (0.05);
\fill[black] (-0.2,0+\y) circle (0.05);
\fill[black] (0.2,0+\y) circle (0.05);
\fill[black] (0.6,0+\y) circle (0.05);
\newcommand{\x}{4}
\draw[thick] (-0.6+\x,0) -- (-0.6+\x,-0.65) arc (180:270:0.15) -- (0.45+\x+3,-0.8) arc (-90:0:0.15) -- (0.6+\x+3,0);
\draw[thick] (-0.2+\x,0) -- (-0.2+\x,-0.45) arc (180:270:0.15) -- (0.05+\x+3,-0.6) arc (-90:0:0.15) -- (0.2+\x+3,0);
\draw[thick] (0.2+\x,0) -- (0.2+\x,-0.25) arc (180:270:0.15) -- (-0.35+\x+3,-0.4) arc (-90:0:0.15) -- (-0.2+\x+3,0);
\draw[thick] (0.6+\x,0) -- (0.6+\x,-0.05) arc (180:270:0.15) -- (-0.75+\x+3,-0.2) arc (-90:0:0.15) -- (-0.6+\x+3,0);
\fill[gray] (0+\x,0) ellipse (0.9cm and 0.2cm);
\fill[black] (-0.6+\x,0) circle (0.05);
\fill[black] (-0.2+\x,0) circle (0.05);
\fill[black] (0.2+\x,0) circle (0.05);
\fill[black] (0.6+\x,0) circle (0.05);
\draw[dashed] (-0.6+\x,0) -- (-0.6+\x,-0.2);
\draw[dashed] (-0.2+\x,0) -- (-0.2+\x,-0.2);
\draw[dashed] (0.2+\x,0) -- (0.2+\x,-0.2);
\draw[dashed] (0.6+\x,0) -- (0.6+\x,-0.05) arc (180:270:0.15);
\fill[gray] (0+\x,0+\y) ellipse (0.9cm and 0.2cm);
\fill[black] (-0.6+\x,0+\y) circle (0.05);
\fill[black] (-0.2+\x,0+\y) circle (0.05);
\fill[black] (0.2+\x,0+\y) circle (0.05);
\fill[black] (0.6+\x,0+\y) circle (0.05);
\draw[thick] (-0.6,0+\y) -- (-0.6,0.75+\y) arc (180:90:0.15) -- (0.45+\x,0.9+\y) arc (90:0:0.15) -- (0.6+\x,0+\y);
\draw[thick] (-0.2,0+\y) -- (-0.2,0.55+\y) arc (180:90:0.15) -- (0.05+\x,0.7+\y) arc (90:0:0.15) -- (0.2+\x,0+\y);
\draw[thick] (0.2,0+\y) -- (0.2,0.35+\y) arc (180:90:0.15) -- (-0.35+\x,0.5+\y) arc (90:0:0.15) -- (-0.2+\x,0+\y);
\draw[thick] (0.6,0+\y) -- (0.6,0.15+\y) arc (180:90:0.15) -- (-0.75+\x,0.3+\y) arc (90:0:0.15) -- (-0.6+\x,0+\y);
\renewcommand{\x}{7}
\fill[gray] (0+\x,0) ellipse (0.9cm and 0.2cm);
\fill[black] (-0.6+\x,0) circle (0.05);
\fill[black] (-0.2+\x,0) circle (0.05);
\fill[black] (0.2+\x,0) circle (0.05);
\fill[black] (0.6+\x,0) circle (0.05);
\draw[dashed] (-0.6+\x,0) -- (-0.6+\x,-0.05) arc (0:-90:0.15);
\draw[dashed] (-0.2+\x,0) -- (-0.2+\x,-0.2);
\draw[dashed] (0.2+\x,0) -- (0.2+\x,-0.2);
\draw[dashed] (0.6+\x,0) -- (0.6+\x,-0.2);
\renewcommand{\y}{0.3}
\draw[blue] (\x-1,0.2+\y) -- (\x-1,0.3+\y) -- (\x+1,0.3+\y) -- (\x+1,0.2+\y);
\draw[blue] (\x,0.5+\y) node {Bob};
\end{tikzpicture}
\end{array}
\qquad \longrightarrow \qquad
\begin{array}{c}
\begin{tikzpicture}
\draw[thick] (-0.6,0) -- (-0.6,-0.8);
\draw[thick] (-0.2,0) -- (-0.2,-0.8);
\draw[thick] (0.2,0) -- (0.2,-0.8);
\draw[thick] (0.6,0) -- (0.6,-0.8);
\fill[gray] (0,0) ellipse (0.9cm and 0.2cm);
\fill[black] (-0.6,0) circle (0.05);
\fill[black] (-0.2,0) circle (0.05);
\fill[black] (0.2,0) circle (0.05);
\fill[black] (0.6,0) circle (0.05);
\draw[dashed] (-0.6,0) -- (-0.6,-0.8);
\draw[dashed] (-0.2,0) -- (-0.2,-0.8);
\draw[dashed] (0.2,0) -- (0.2,-0.8);
\draw[dashed] (0.6,0) -- (0.6,-0.8);
\fill[black] (0,-0.8) ellipse (0.9cm and 0.2cm);
\newcommand{\x}{0}
\draw[blue] (\x-1,0.2) -- (\x-1,0.3) -- (\x+1,0.3) -- (\x+1,0.2);
\draw[blue] (\x,0.5) node {Bob};
\end{tikzpicture}
\end{array}
\ee
In this diagram one glues three 3-manifolds along a pair of boundaries -- the inner product operation in TQFT. As a result, Bob trivially obtains the original state by isotopy (by pulling the strings). However, the operation in~(\ref{projection1}) is a projection on a specific state, in fact, the same state as the initial pair of entangled qubits in~(\ref{pairofstates}), so teleporting a state this way is only possible with a finite probability. Although this will be sufficient for our later discussion, let us first discuss more practical versions of the teleportation.

In order to achieve the teleportation with certainty Alice can use complete orthonormal basis of entangled vectors on the pair of qubits. State projected upon in~(\ref{projection1}) is equivalent to the Bell state $|00\rangle+|11\rangle$ (see~\cite{Melnikov:2022qyt}). Alice can make measurements in the Bell basis of entangled pairs, 
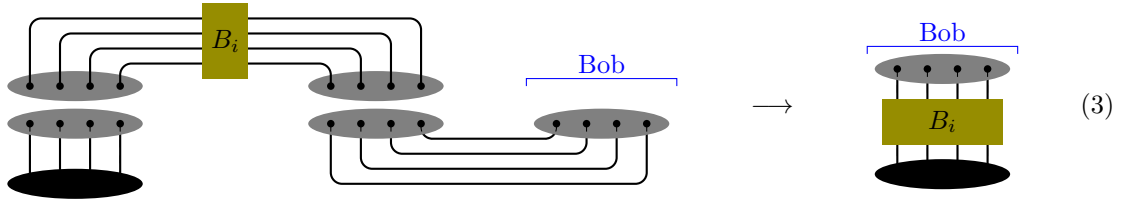
\be
\label{projection2}
\begin{array}{c}
\begin{tikzpicture}
\draw[thick] (-0.6,0) -- (-0.6,-0.8);
\draw[thick] (-0.2,0) -- (-0.2,-0.8);
\draw[thick] (0.2,0) -- (0.2,-0.8);
\draw[thick] (0.6,0) -- (0.6,-0.8);
\fill[gray] (0,0) ellipse (0.9cm and 0.2cm);
\fill[black] (-0.6,0) circle (0.05);
\fill[black] (-0.2,0) circle (0.05);
\fill[black] (0.2,0) circle (0.05);
\fill[black] (0.6,0) circle (0.05);
\draw[dashed] (-0.6,0) -- (-0.6,-0.8);
\draw[dashed] (-0.2,0) -- (-0.2,-0.8);
\draw[dashed] (0.2,0) -- (0.2,-0.8);
\draw[dashed] (0.6,0) -- (0.6,-0.8);
\fill[black] (0,-0.8) ellipse (0.9cm and 0.2cm);
\newcommand{\y}{0.5}
\fill[gray] (0,0+\y) ellipse (0.9cm and 0.2cm);
\fill[black] (-0.6,0+\y) circle (0.05);
\fill[black] (-0.2,0+\y) circle (0.05);
\fill[black] (0.2,0+\y) circle (0.05);
\fill[black] (0.6,0+\y) circle (0.05);
\newcommand{\x}{4}
\draw[thick] (-0.6+\x,0) -- (-0.6+\x,-0.65) arc (180:270:0.15) -- (0.45+\x+3,-0.8) arc (-90:0:0.15) -- (0.6+\x+3,0);
\draw[thick] (-0.2+\x,0) -- (-0.2+\x,-0.45) arc (180:270:0.15) -- (0.05+\x+3,-0.6) arc (-90:0:0.15) -- (0.2+\x+3,0);
\draw[thick] (0.2+\x,0) -- (0.2+\x,-0.25) arc (180:270:0.15) -- (-0.35+\x+3,-0.4) arc (-90:0:0.15) -- (-0.2+\x+3,0);
\draw[thick] (0.6+\x,0) -- (0.6+\x,-0.05) arc (180:270:0.15) -- (-0.75+\x+3,-0.2) arc (-90:0:0.15) -- (-0.6+\x+3,0);
\fill[gray] (0+\x,0) ellipse (0.9cm and 0.2cm);
\fill[black] (-0.6+\x,0) circle (0.05);
\fill[black] (-0.2+\x,0) circle (0.05);
\fill[black] (0.2+\x,0) circle (0.05);
\fill[black] (0.6+\x,0) circle (0.05);
\draw[dashed] (-0.6+\x,0) -- (-0.6+\x,-0.2);
\draw[dashed] (-0.2+\x,0) -- (-0.2+\x,-0.2);
\draw[dashed] (0.2+\x,0) -- (0.2+\x,-0.2);
\draw[dashed] (0.6+\x,0) -- (0.6+\x,-0.05) arc (180:270:0.15);
\fill[gray] (0+\x,0+\y) ellipse (0.9cm and 0.2cm);
\fill[black] (-0.6+\x,0+\y) circle (0.05);
\fill[black] (-0.2+\x,0+\y) circle (0.05);
\fill[black] (0.2+\x,0+\y) circle (0.05);
\fill[black] (0.6+\x,0+\y) circle (0.05);
\draw[thick] (-0.6,0+\y) -- (-0.6,0.75+\y) arc (180:90:0.15) -- (0.45+\x,0.9+\y) arc (90:0:0.15) -- (0.6+\x,0+\y);
\draw[thick] (-0.2,0+\y) -- (-0.2,0.55+\y) arc (180:90:0.15) -- (0.05+\x,0.7+\y) arc (90:0:0.15) -- (0.2+\x,0+\y);
\draw[thick] (0.2,0+\y) -- (0.2,0.35+\y) arc (180:90:0.15) -- (-0.35+\x,0.5+\y) arc (90:0:0.15) -- (-0.2+\x,0+\y);
\draw[thick] (0.6,0+\y) -- (0.6,0.15+\y) arc (180:90:0.15) -- (-0.75+\x,0.3+\y) arc (90:0:0.15) -- (-0.6+\x,0+\y);
\fill[olive] (\x/2-0.3,\y+0.1) rectangle (\x/2+0.3,\y+1.1); 
\draw (\x/2,\y+0.6) node {$B_i$};
\renewcommand{\x}{7}
\fill[gray] (0+\x,0) ellipse (0.9cm and 0.2cm);
\fill[black] (-0.6+\x,0) circle (0.05);
\fill[black] (-0.2+\x,0) circle (0.05);
\fill[black] (0.2+\x,0) circle (0.05);
\fill[black] (0.6+\x,0) circle (0.05);
\draw[dashed] (-0.6+\x,0) -- (-0.6+\x,-0.05) arc (0:-90:0.15);
\draw[dashed] (-0.2+\x,0) -- (-0.2+\x,-0.2);
\draw[dashed] (0.2+\x,0) -- (0.2+\x,-0.2);
\draw[dashed] (0.6+\x,0) -- (0.6+\x,-0.2);
\renewcommand{\y}{0.3}
\draw[blue] (\x-1,0.2+\y) -- (\x-1,0.3+\y) -- (\x+1,0.3+\y) -- (\x+1,0.2+\y);
\draw[blue] (\x,0.5+\y) node {Bob};
\end{tikzpicture}
\end{array}
\qquad \longrightarrow \qquad
\begin{array}{c}
\begin{tikzpicture}
\newcommand{\y}{0.6}
\draw[thick] (-0.6,0) -- (-0.6,-0.8-\y);
\draw[thick] (-0.2,0) -- (-0.2,-0.8-\y);
\draw[thick] (0.2,0) -- (0.2,-0.8-\y);
\draw[thick] (0.6,0) -- (0.6,-0.8-\y);
\fill[gray] (0,0) ellipse (0.9cm and 0.2cm);
\fill[black] (-0.6,0) circle (0.05);
\fill[black] (-0.2,0) circle (0.05);
\fill[black] (0.2,0) circle (0.05);
\fill[black] (0.6,0) circle (0.05);
\draw[dashed] (-0.6,0) -- (-0.6,-0.8-\y);
\draw[dashed] (-0.2,0) -- (-0.2,-0.8-\y);
\draw[dashed] (0.2,0) -- (0.2,-0.8-\y);
\draw[dashed] (0.6,0) -- (0.6,-0.8-\y);
\fill[black] (0,-0.8-\y) ellipse (0.9cm and 0.2cm);
\fill[olive] (-0.8,-\y/2-0.7) rectangle (0.8,-\y/2-0.1); 
\draw (0,-\y/2-0.4) node {$B_i$};
\newcommand{\x}{0}
\renewcommand{\y}{0}
\draw[blue] (\x-1,\y+0.2) -- (\x-1,\y+0.3) -- (\x+1,\y+0.3) -- (\x+1,\y+0.2);
\draw[blue] (\x,\y+0.5) node {Bob};
\end{tikzpicture}
\end{array}
\ee
where $B_i$ label the vectors of the Bell basis, or can represent any orthonormal set of entangled states. Entanglement guaranties that states $B_i$ are equivalent to invertible operators acting on the teleported qubit, as the right part of~(\ref{projection2}) implies. It is left to Bob to apply the inverse operation to undo the effects of $B_i$. This is the version of the protocol discussed in the context of cathegorical quantum mechanics~\cite{Coecke:2006qua,Coecke} based on Penrose diagrammatical notation~\cite{Penrose}.

In the TQFT realization one may be interested in defining $B_i$ as natural topological operations, like braiding. The problem however, that there are no naturally orthogonal topological presentation of states. At most appropriate orthogonal $B_i$ can be constructed as linear combinations of the diagrams. If one insists on using natural operations, without introducing boxes, like in equation~(\ref{projection2}), the projection will have to be done on nonorthogonal basis, as in the following construction.

As in the standard teleportation protocol~\cite{Bennett:1992tv,NielsenChuang} one first entangles the state that is being teleported with the Alice's copy of the entangled pair. In TQFT this can be achieved through braiding, which in general is an entangling operation~\cite{Kauffman}. Let us consider two options:
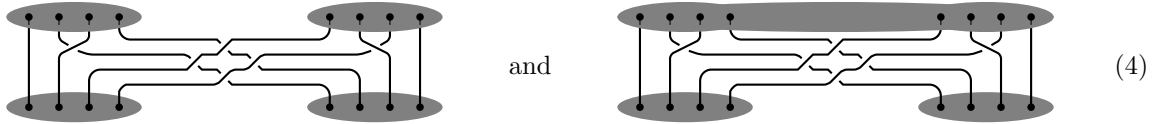
\be
\label{entangler}
\begin{array}{c}
\begin{tikzpicture}
\newcommand{\y}{0.5}
\fill[gray] (0,0+\y) ellipse (0.9cm and 0.2cm);
\fill[black] (-0.6,0+\y) circle (0.05);
\fill[black] (-0.2,0+\y) circle (0.05);
\fill[black] (0.2,0+\y) circle (0.05);
\fill[black] (0.6,0+\y) circle (0.05);
\newcommand{\x}{4}
\fill[gray] (0+\x,0+\y) ellipse (0.9cm and 0.2cm);
\fill[black] (-0.6+\x,0+\y) circle (0.05);
\fill[black] (-0.2+\x,0+\y) circle (0.05);
\fill[black] (0.2+\x,0+\y) circle (0.05);
\fill[black] (0.6+\x,0+\y) circle (0.05);
\draw[thick] (-0.6,0+\y) -- (-0.6,1.2+\y);
\draw[thick] (0.6+\x,0+\y) -- (0.6+\x,1.2+\y);
\draw[thick,rounded corners=1.5] (-0.2,0+\y) -- (-0.2,0.7+\y) -- (0.2,0.9+\y) -- (0.2,1.2+\y);
\draw[thick,rounded corners=1.5] (0.2+\x,0+\y) -- (0.2+\x,0.7+\y) -- (-0.2+\x,0.9+\y) -- (-0.2+\x,1.2+\y);
\draw[thick] (0.2,0+\y) -- (0.2,0.35+\y) arc (180:90:0.15) -- (\x/2-0.5,0.5+\y) -- (\x/2-0.3,0.7+\y)-- (\x/2-0.1,0.7+\y) -- (\x/2+0.1,0.9+\y) -- (\x-0.75,0.9+\y) arc (-90:0:0.15) -- (-0.6+\x,1.2+\y);
\draw[thick,rounded corners=1.5] (0.6,0+\y) -- (0.6,0.15+\y) arc (180:90:0.15) -- (\x/2-0.1,0.3+\y) -- (\x/2+0.1,0.5+\y) -- (\x/2+0.3,0.5+\y) -- (\x/2+0.5,0.7+\y) -- (\x-0.2,0.7+\y) -- (\x-0.05,0.75+\y);
\draw[thick,rounded corners=1.5] (\x+0.0725,0.825+\y) -- (\x+0.2,0.9+\y) -- (\x+0.2,1.2+\y);
\draw[thick] (-0.6+\x,0+\y) -- (-0.6+\x,0.15+\y) arc (0:90:0.15) -- (\x/2+0.1,0.3+\y) -- (\x/2+0.05,0.35+\y);
\draw[thick] (\x/2-0.05,0.45+\y) -- (\x/2-0.1,0.5+\y) -- (\x/2-0.3,0.5+\y) -- (\x/2-0.35,0.55+\y);
\draw[thick,rounded corners=1.5] (\x/2-0.45,0.65+\y) -- (\x/2-0.5,0.7+\y) -- (0.2,0.7+\y) -- (0.05,0.75+\y);
\draw[thick,rounded corners=1.5] (-0.075,0.825+\y) -- (-0.2,0.9+\y) -- (-0.2,1.2+\y);
\draw[thick] (-0.2+\x,0+\y) -- (-0.2+\x,0.35+\y) arc (0:90:0.15) -- (\x/2+0.5,0.5+\y) -- (\x/2+0.45,0.55+\y);
\draw[thick] (\x/2+0.35,0.65+\y) -- (\x/2+0.3,0.7+\y) -- (\x/2+0.1,0.7+\y) -- (\x/2+0.05,0.75+\y);
\draw[thick] (\x/2-0.05,0.85+\y) -- (\x/2-0.1,0.9+\y) -- (0.75,0.9+\y) arc (-90:-180:0.15) -- (0.6,1.2+\y);
\renewcommand{\y}{1.7}
\fill[gray] (0,0+\y) ellipse (0.9cm and 0.2cm);
\fill[black] (-0.6,0+\y) circle (0.05);
\fill[black] (-0.2,0+\y) circle (0.05);
\fill[black] (0.2,0+\y) circle (0.05);
\fill[black] (0.6,0+\y) circle (0.05);
\fill[gray] (0+\x,0+\y) ellipse (0.9cm and 0.2cm);
\fill[black] (-0.6+\x,0+\y) circle (0.05);
\fill[black] (-0.2+\x,0+\y) circle (0.05);
\fill[black] (0.2+\x,0+\y) circle (0.05);
\fill[black] (0.6+\x,0+\y) circle (0.05);
\draw[dashed] (-0.6+\x,0+\y) -- (-0.6+\x,-0.15+\y) arc (0:-90:0.15);
\draw[dashed] (-0.2+\x,0+\y) -- (-0.2+\x,-0.2+\y);
\draw[dashed] (0.2+\x,0+\y) -- (0.2+\x,-0.2+\y);
\draw[dashed] (0.6+\x,0+\y) -- (0.6+\x,-0.2+\y);
\draw[dashed] (-0.6,0+\y) -- (-0.6,-0.15+\y);
\draw[dashed] (-0.2,0+\y) -- (-0.2,-0.2+\y);
\draw[dashed] (0.2,0+\y) -- (0.2,-0.2+\y);
\draw[dashed] (0.6,0+\y) -- (0.6,-0.15+\y) arc (180:270:0.15);
\end{tikzpicture}
\end{array}
\qquad \text{and} \qquad
\begin{array}{c}
\begin{tikzpicture}
\newcommand{\y}{0.5}
\fill[gray] (0,0+\y) ellipse (0.9cm and 0.2cm);
\fill[black] (-0.6,0+\y) circle (0.05);
\fill[black] (-0.2,0+\y) circle (0.05);
\fill[black] (0.2,0+\y) circle (0.05);
\fill[black] (0.6,0+\y) circle (0.05);
\newcommand{\x}{4}
\fill[gray] (0+\x,0+\y) ellipse (0.9cm and 0.2cm);
\fill[black] (-0.6+\x,0+\y) circle (0.05);
\fill[black] (-0.2+\x,0+\y) circle (0.05);
\fill[black] (0.2+\x,0+\y) circle (0.05);
\fill[black] (0.6+\x,0+\y) circle (0.05);
\draw[thick] (-0.6,0+\y) -- (-0.6,1.2+\y);
\draw[thick] (0.6+\x,0+\y) -- (0.6+\x,1.2+\y);
\draw[thick,rounded corners=1.5] (-0.2,0+\y) -- (-0.2,0.7+\y) -- (0.2,0.9+\y) -- (0.2,1.2+\y);
\draw[thick,rounded corners=1.5] (0.2+\x,0+\y) -- (0.2+\x,0.7+\y) -- (-0.2+\x,0.9+\y) -- (-0.2+\x,1.2+\y);
\draw[thick] (0.2,0+\y) -- (0.2,0.35+\y) arc (180:90:0.15) -- (\x/2-0.5,0.5+\y) -- (\x/2-0.3,0.7+\y)-- (\x/2-0.1,0.7+\y) -- (\x/2+0.1,0.9+\y) -- (\x-0.75,0.9+\y) arc (-90:0:0.15) -- (-0.6+\x,1.2+\y);
\draw[thick,rounded corners=1.5] (0.6,0+\y) -- (0.6,0.15+\y) arc (180:90:0.15) -- (\x/2-0.1,0.3+\y) -- (\x/2+0.1,0.5+\y) -- (\x/2+0.3,0.5+\y) -- (\x/2+0.5,0.7+\y) -- (\x-0.2,0.7+\y) -- (\x-0.05,0.75+\y);
\draw[thick,rounded corners=1.5] (\x+0.0725,0.825+\y) -- (\x+0.2,0.9+\y) -- (\x+0.2,1.2+\y);
\draw[thick] (-0.6+\x,0+\y) -- (-0.6+\x,0.15+\y) arc (0:90:0.15) -- (\x/2+0.1,0.3+\y) -- (\x/2+0.05,0.35+\y);
\draw[thick] (\x/2-0.05,0.45+\y) -- (\x/2-0.1,0.5+\y) -- (\x/2-0.3,0.5+\y) -- (\x/2-0.35,0.55+\y);
\draw[thick,rounded corners=1.5] (\x/2-0.45,0.65+\y) -- (\x/2-0.5,0.7+\y) -- (0.2,0.7+\y) -- (0.05,0.75+\y);
\draw[thick,rounded corners=1.5] (-0.075,0.825+\y) -- (-0.2,0.9+\y) -- (-0.2,1.2+\y);
\draw[thick] (-0.2+\x,0+\y) -- (-0.2+\x,0.35+\y) arc (0:90:0.15) -- (\x/2+0.5,0.5+\y) -- (\x/2+0.45,0.55+\y);
\draw[thick] (\x/2+0.35,0.65+\y) -- (\x/2+0.3,0.7+\y) -- (\x/2+0.1,0.7+\y) -- (\x/2+0.05,0.75+\y);
\draw[thick] (\x/2-0.05,0.85+\y) -- (\x/2-0.1,0.9+\y) -- (0.75,0.9+\y) arc (-90:-180:0.15) -- (0.6,1.2+\y);
\renewcommand{\y}{1.7}
\fill[gray] (0,0+\y) ellipse (0.9cm and 0.2cm);
\fill[gray] (0+\x/2,0+\y) ellipse (2.9cm and 0.2cm);
\fill[black] (-0.6,0+\y) circle (0.05);
\fill[black] (-0.2,0+\y) circle (0.05);
\fill[black] (0.2,0+\y) circle (0.05);
\fill[black] (0.6,0+\y) circle (0.05);
\fill[gray] (0+\x,0+\y) ellipse (0.9cm and 0.2cm);
\fill[black] (-0.6+\x,0+\y) circle (0.05);
\fill[black] (-0.2+\x,0+\y) circle (0.05);
\fill[black] (0.2+\x,0+\y) circle (0.05);
\fill[black] (0.6+\x,0+\y) circle (0.05);
\draw[dashed] (-0.6+\x,0+\y) -- (-0.6+\x,-0.15+\y) arc (0:-90:0.15);
\draw[dashed] (-0.2+\x,0+\y) -- (-0.2+\x,-0.2+\y);
\draw[dashed] (0.2+\x,0+\y) -- (0.2+\x,-0.2+\y);
\draw[dashed] (0.6+\x,0+\y) -- (0.6+\x,-0.2+\y);
\draw[dashed] (-0.6,0+\y) -- (-0.6,-0.15+\y);
\draw[dashed] (-0.2,0+\y) -- (-0.2,-0.2+\y);
\draw[dashed] (0.2,0+\y) -- (0.2,-0.2+\y);
\draw[dashed] (0.6,0+\y) -- (0.6,-0.15+\y) arc (180:270:0.15);
\end{tikzpicture}
\end{array}
\ee
These two diagrams correspond to two kinds of linear maps: the first is an operator on two qubits, while the second is an operator on two qubits embedded in a larger Hilbert space. The first option seems a natural choice, but in the TQFT context the corresponding operator is in general not unitary. It can be made unitary by embedding it in a larger space, as the second diagram suggests.\footnote{In the specific setup of $SU(2)$ Chern-Simons theory the dimension of the Hilbert space associated with a boundary depends on the number of lines that end on it.}

After the application of~(\ref{entangler}) Alice can measure the result, projecting the entangled state on an appropriate basis. A convenient TQFT basis is provided by a set of diagrams that connect points on the same boundary by lines that do not intersect, that is one can pull the strings to make them nonintersecting in the given projection on a plane. One example is the projection on the following state:
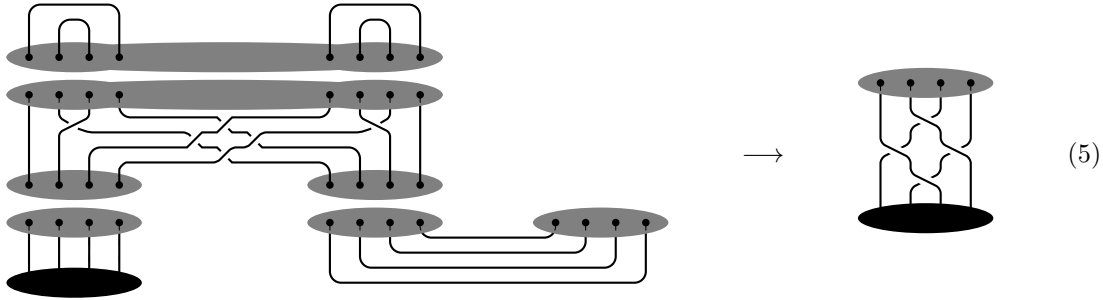
\be
\label{teleportation}
\begin{array}{c}
\begin{tikzpicture}
\draw[thick] (-0.6,0) -- (-0.6,-0.8);
\draw[thick] (-0.2,0) -- (-0.2,-0.8);
\draw[thick] (0.2,0) -- (0.2,-0.8);
\draw[thick] (0.6,0) -- (0.6,-0.8);
\fill[gray] (0,0) ellipse (0.9cm and 0.2cm);
\fill[black] (-0.6,0) circle (0.05);
\fill[black] (-0.2,0) circle (0.05);
\fill[black] (0.2,0) circle (0.05);
\fill[black] (0.6,0) circle (0.05);
\draw[dashed] (-0.6,0) -- (-0.6,-0.8);
\draw[dashed] (-0.2,0) -- (-0.2,-0.8);
\draw[dashed] (0.2,0) -- (0.2,-0.8);
\draw[dashed] (0.6,0) -- (0.6,-0.8);
\fill[black] (0,-0.8) ellipse (0.9cm and 0.2cm);
\newcommand{\y}{0.5}
\fill[gray] (0,0+\y) ellipse (0.9cm and 0.2cm);
\fill[black] (-0.6,0+\y) circle (0.05);
\fill[black] (-0.2,0+\y) circle (0.05);
\fill[black] (0.2,0+\y) circle (0.05);
\fill[black] (0.6,0+\y) circle (0.05);
\newcommand{\x}{4}
\draw[thick] (-0.6+\x,0) -- (-0.6+\x,-0.65) arc (180:270:0.15) -- (0.45+\x+3,-0.8) arc (-90:0:0.15) -- (0.6+\x+3,0);
\draw[thick] (-0.2+\x,0) -- (-0.2+\x,-0.45) arc (180:270:0.15) -- (0.05+\x+3,-0.6) arc (-90:0:0.15) -- (0.2+\x+3,0);
\draw[thick] (0.2+\x,0) -- (0.2+\x,-0.25) arc (180:270:0.15) -- (-0.35+\x+3,-0.4) arc (-90:0:0.15) -- (-0.2+\x+3,0);
\draw[thick] (0.6+\x,0) -- (0.6+\x,-0.05) arc (180:270:0.15) -- (-0.75+\x+3,-0.2) arc (-90:0:0.15) -- (-0.6+\x+3,0);
\fill[gray] (0+\x,0) ellipse (0.9cm and 0.2cm);
\fill[black] (-0.6+\x,0) circle (0.05);
\fill[black] (-0.2+\x,0) circle (0.05);
\fill[black] (0.2+\x,0) circle (0.05);
\fill[black] (0.6+\x,0) circle (0.05);
\draw[dashed] (-0.6+\x,0) -- (-0.6+\x,-0.2);
\draw[dashed] (-0.2+\x,0) -- (-0.2+\x,-0.2);
\draw[dashed] (0.2+\x,0) -- (0.2+\x,-0.2);
\draw[dashed] (0.6+\x,0) -- (0.6+\x,-0.05) arc (180:270:0.15);
\fill[gray] (0+\x,0+\y) ellipse (0.9cm and 0.2cm);
\fill[black] (-0.6+\x,0+\y) circle (0.05);
\fill[black] (-0.2+\x,0+\y) circle (0.05);
\fill[black] (0.2+\x,0+\y) circle (0.05);
\fill[black] (0.6+\x,0+\y) circle (0.05);
\draw[thick] (-0.6,0+\y) -- (-0.6,1.2+\y);
\draw[thick] (0.6+\x,0+\y) -- (0.6+\x,1.2+\y);
\draw[thick,rounded corners=1.5] (-0.2,0+\y) -- (-0.2,0.7+\y) -- (0.2,0.9+\y) -- (0.2,1.2+\y);
\draw[thick,rounded corners=1.5] (0.2+\x,0+\y) -- (0.2+\x,0.7+\y) -- (-0.2+\x,0.9+\y) -- (-0.2+\x,1.2+\y);
\draw[thick] (0.2,0+\y) -- (0.2,0.35+\y) arc (180:90:0.15) -- (\x/2-0.5,0.5+\y) -- (\x/2-0.3,0.7+\y)-- (\x/2-0.1,0.7+\y) -- (\x/2+0.1,0.9+\y) -- (\x-0.75,0.9+\y) arc (-90:0:0.15) -- (-0.6+\x,1.2+\y);
\draw[thick,rounded corners=1.5] (0.6,0+\y) -- (0.6,0.15+\y) arc (180:90:0.15) -- (\x/2-0.1,0.3+\y) -- (\x/2+0.1,0.5+\y) -- (\x/2+0.3,0.5+\y) -- (\x/2+0.5,0.7+\y) -- (\x-0.2,0.7+\y) -- (\x-0.05,0.75+\y);
\draw[thick,rounded corners=1.5] (\x+0.0725,0.825+\y) -- (\x+0.2,0.9+\y) -- (\x+0.2,1.2+\y);
\draw[thick] (-0.6+\x,0+\y) -- (-0.6+\x,0.15+\y) arc (0:90:0.15) -- (\x/2+0.1,0.3+\y) -- (\x/2+0.05,0.35+\y);
\draw[thick] (\x/2-0.05,0.45+\y) -- (\x/2-0.1,0.5+\y) -- (\x/2-0.3,0.5+\y) -- (\x/2-0.35,0.55+\y);
\draw[thick,rounded corners=1.5] (\x/2-0.45,0.65+\y) -- (\x/2-0.5,0.7+\y) -- (0.2,0.7+\y) -- (0.05,0.75+\y);
\draw[thick,rounded corners=1.5] (-0.075,0.825+\y) -- (-0.2,0.9+\y) -- (-0.2,1.2+\y);
\draw[thick] (-0.2+\x,0+\y) -- (-0.2+\x,0.35+\y) arc (0:90:0.15) -- (\x/2+0.5,0.5+\y) -- (\x/2+0.45,0.55+\y);
\draw[thick] (\x/2+0.35,0.65+\y) -- (\x/2+0.3,0.7+\y) -- (\x/2+0.1,0.7+\y) -- (\x/2+0.05,0.75+\y);
\draw[thick] (\x/2-0.05,0.85+\y) -- (\x/2-0.1,0.9+\y) -- (0.75,0.9+\y) arc (-90:-180:0.15) -- (0.6,1.2+\y);
\renewcommand{\y}{1.7}
\fill[gray] (0,0+\y) ellipse (0.9cm and 0.2cm);
\fill[gray] (0+\x/2,0+\y) ellipse (2.9cm and 0.2cm);
\fill[black] (-0.6,0+\y) circle (0.05);
\fill[black] (-0.2,0+\y) circle (0.05);
\fill[black] (0.2,0+\y) circle (0.05);
\fill[black] (0.6,0+\y) circle (0.05);
\fill[gray] (0+\x,0+\y) ellipse (0.9cm and 0.2cm);
\fill[black] (-0.6+\x,0+\y) circle (0.05);
\fill[black] (-0.2+\x,0+\y) circle (0.05);
\fill[black] (0.2+\x,0+\y) circle (0.05);
\fill[black] (0.6+\x,0+\y) circle (0.05);
\draw[dashed] (-0.6+\x,0+\y) -- (-0.6+\x,-0.15+\y) arc (0:-90:0.15);
\draw[dashed] (-0.2+\x,0+\y) -- (-0.2+\x,-0.2+\y);
\draw[dashed] (0.2+\x,0+\y) -- (0.2+\x,-0.2+\y);
\draw[dashed] (0.6+\x,0+\y) -- (0.6+\x,-0.2+\y);
\draw[dashed] (-0.6,0+\y) -- (-0.6,-0.15+\y);
\draw[dashed] (-0.2,0+\y) -- (-0.2,-0.2+\y);
\draw[dashed] (0.2,0+\y) -- (0.2,-0.2+\y);
\draw[dashed] (0.6,0+\y) -- (0.6,-0.15+\y) arc (180:270:0.15);
\renewcommand{\y}{2.2}
\fill[gray] (0,0+\y) ellipse (0.9cm and 0.2cm);
\fill[gray] (0+\x/2,0+\y) ellipse (2.9cm and 0.2cm);
\fill[black] (-0.6,0+\y) circle (0.05);
\fill[black] (-0.2,0+\y) circle (0.05);
\fill[black] (0.2,0+\y) circle (0.05);
\fill[black] (0.6,0+\y) circle (0.05);
\fill[gray] (0+\x,0+\y) ellipse (0.9cm and 0.2cm);
\fill[black] (-0.6+\x,0+\y) circle (0.05);
\fill[black] (-0.2+\x,0+\y) circle (0.05);
\fill[black] (0.2+\x,0+\y) circle (0.05);
\fill[black] (0.6+\x,0+\y) circle (0.05);
\draw[thick] (-0.2,0+\y) -- (-0.2,0.35+\y) arc (180:90:0.15) -- (0.05,0.5+\y) arc (90:0:0.15) -- (0.2,0+\y);
\draw[thick] (-0.6,0+\y) -- (-0.6,0.55+\y) arc (180:90:0.15) -- (0.45,0.7+\y) arc (90:0:0.15) -- (0.6,0+\y);
\draw[thick] (-0.2+\x,0+\y) -- (-0.2+\x,0.35+\y) arc (180:90:0.15) -- (0.05+\x,0.5+\y) arc (90:0:0.15) -- (0.2+\x,0+\y);
\draw[thick] (-0.6+\x,0+\y) -- (-0.6+\x,0.55+\y) arc (180:90:0.15) -- (0.45+\x,0.7+\y) arc (90:0:0.15) -- (0.6+\x,0+\y);
\renewcommand{\x}{7}
\fill[gray] (0+\x,0) ellipse (0.9cm and 0.2cm);
\fill[black] (-0.6+\x,0) circle (0.05);
\fill[black] (-0.2+\x,0) circle (0.05);
\fill[black] (0.2+\x,0) circle (0.05);
\fill[black] (0.6+\x,0) circle (0.05);
\draw[dashed] (-0.6+\x,0) -- (-0.6+\x,-0.05) arc (0:-90:0.15);
\draw[dashed] (-0.2+\x,0) -- (-0.2+\x,-0.2);
\draw[dashed] (0.2+\x,0) -- (0.2+\x,-0.2);
\draw[dashed] (0.6+\x,0) -- (0.6+\x,-0.2);
\renewcommand{\y}{0.3}
\end{tikzpicture}
\end{array}
\qquad \longrightarrow \qquad
\begin{array}{c}
\begin{tikzpicture}{thick}
\draw[thick,rounded corners=2] (0.2,0) -- (0.2,-0.4) -- (-0.2,-0.6) -- (-0.2,-0.8) -- (-0.6,-1.0) -- (-0.6,-1.8);
\draw[thick,rounded corners=2] (0.6,0) -- (0.6,-0.8) -- (0.2,-1.0) -- (0.2,-1.2) -- (-0.2,-1.4) -- (-0.2,-1.8);
\draw[line width=4,white,rounded corners=2] (-0.6,0) -- (-0.6,-0.8) -- (-0.2,-1.0) -- (-0.2, -1.2) -- (0.2,-1.4) -- (0.2,-1.8);
\draw[line width=4,white,rounded corners=2] (-0.2,0) -- (-0.2,-0.4) -- (0.2, -0.6) -- (0.2,-0.8) -- (0.6,-1.0) -- (0.6,-1.8);
\draw[thick,rounded corners=2] (-0.6,0) -- (-0.6,-0.8) -- (-0.2,-1.0) -- (-0.2, -1.2) -- (0.2,-1.4) -- (0.2,-1.8);
\draw[thick,rounded corners=2] (-0.2,0) -- (-0.2,-0.4) -- (0.2, -0.6) -- (0.2,-0.8) -- (0.6,-1.0) -- (0.6,-1.8);
\fill[gray] (0,0) ellipse (0.9cm and 0.2cm);
\fill[black] (-0.6,0) circle (0.05);
\fill[black] (-0.2,0) circle (0.05);
\fill[black] (0.2,0) circle (0.05);
\fill[black] (0.6,0) circle (0.05);
\draw[dashed] (-0.6,0) -- (-0.6,-0.2);
\draw[dashed] (-0.2,0) -- (-0.2,-0.2);
\draw[dashed] (0.2,0) -- (0.2,-0.2);
\draw[dashed] (0.6,0) -- (0.6,-0.2);
\fill[black] (0,-1.8) ellipse (0.9cm and 0.2cm);
\end{tikzpicture}
\end{array}
\ee
The projection is simply the gluing of a basis state to the open boundary corresponding to Alice's Hilbert space. It is easy to see that the result of this operation is the topology shown on the right. As before Bob still needs to apply some local braiding operation on his side. Which precise braiding is necessary will be communicated to him by Alice, pending the result of her measurement. One can check that for all fourteen possible basis vectors the result will be similar to that in~(\ref{teleportation}), and the teleported state can be recovered through a local braiding in Bob's space. 

Protocol~(\ref{teleportation}) is, of course, a particular case of~(\ref{projection2}), with $B_i$ being a set of nonorthogonal braiding operators. As such the teleportation will have finite fidelity. Nonetheless, orthogonality of braiding operations can be achieved in specific TQFT realizations, such as Chern-Simons theories with adjusted level.

\bigskip

We now turn to an application of the teleportation protocol. The present discussion contains no notion of continuous distance, but one can introduce a discrete distance by declaring that Alice's qubits are local, while Bob's one is in a causally disconnected region. In other words, one can assume that the qubits occupy positions on a lattice, with some lattice sites belonging to Alice and the others to Bob, and that there is a barrier between the two sets. For example, Alice can be in the interior of a black hole, but with a classical communication channel with Bob, who is outside of the black hole's horizon. Despite causal separation, lines expressing quantum correlations, can connect any two parts of the system.

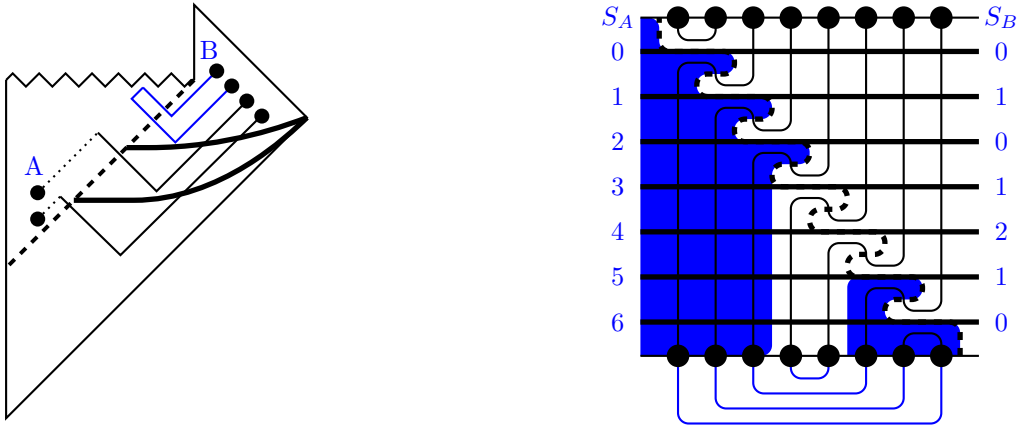
\begin{figure}[t]
\begin{minipage}{0.45\linewidth}
    \centering
    \begin{tikzpicture}
    \draw[thick] (0,4.5) -- (0,0) -- (4,4) -- (2.5,5.5) -- (2.5,4.5);
    \draw[thick,decorate,decoration=zigzag] (0,4.5) -- (2.5,4.5);
    \draw[thick,dashed,line width=1.5] (2.5,4.5) -- (0,2);
    \fill[black] (0.42,3) circle (0.1);
    \fill[black] (3.2,4.22) circle (0.1); 
    \draw[thick,dotted] (0.42,3) -- (1.22,3.8);
    \draw[thick] (1.22,3.8) -- (2.0,3.02) -- (3.2,4.22);
    \fill[black] (0.42,2.65) circle (0.1);
    \draw[thick,dotted] (0.42,2.65) -- (0.72,2.95);
    \draw[thick] (0.72,2.95) -- (1.52,2.17) -- (3.4,4.03);
    \fill[black] (3.4,4.02) circle (0.1);
    \draw[thick,blue] (2.8,4.62) -- (2.2,4.02) -- (1.82,4.4);
    \draw[thick,blue] (3.0,4.42) -- (2.25,3.67) -- (1.67,4.25) -- (1.82,4.4);
    \fill[black] (3.0,4.42) circle (0.1);
    \fill[black] (2.8,4.62) circle (0.1);
    \draw[blue] (2.7,4.9) node {B};
    \draw[blue] (0.37,3.35) node {A};
    \draw[thick,line width=2,opacity=0.3] (1.6,3.6) parabola bend (1.9,3.6) (4,4);
    \draw[thick,line width=2,opacity=0.3] (0.9,2.9) parabola bend (1.7,2.9) (4,4);
    \end{tikzpicture}
    \end{minipage}
    \hfill{
    \begin{minipage}{0.45\linewidth}
        \centering
        \begin{tikzpicture}
        \fill[blue,rounded corners=4,opacity=0.1] (4.25,0) -- (4.25,0.45) -- (3.25,0.45) -- (3.25,0.75) -- (3.75,0.75) -- (3.75,1.05) -- (2.75,1.05) -- (2.75,0.0);
        \fill[blue,rounded corners=4,opacity=0.1] (0,0) -- (1.75,0) -- (1.75,2.55) -- (2.25,2.55) -- (2.25,2.85) -- (1.25,2.85) -- (1.25,3.15) -- (1.75,3.15) -- (1.75,3.45) -- (0.75,3.45) -- (0.75,3.75) -- (1.25,3.75) -- (1.25,4.05) -- (0.25,4.05) -- (0.25,4.5) -- (0,4.5);
        \foreach \a in {0.45,1.05,...,4.65}
            \draw[line width=2,opacity=0.1] (0,\a) -- (4.5,\a);
        \draw[thick] (0,0) -- (4.5,0);
        \draw[thick] (0,4.5) -- (4.5,4.5);
        \draw[blue,thick,rounded corners=4] (2,0) -- (2,-0.3) -- (2.5,-0.3) -- (2.5,0);
        \draw[blue,thick,rounded corners=4] (1.5,0) -- (1.5,-0.5) -- (3.0,-0.5) -- (3.0,0);
        \draw[blue,thick,rounded corners=4] (1.,0) -- (1.,-0.7) -- (3.5,-0.7) -- (3.5,0);
        \draw[blue,thick,rounded corners=4] (0.5,0) -- (0.5,-0.9) -- (4.0,-0.9) -- (4.0,0);
        \foreach \a in {0.5,1.0,...,4.0}
            \fill[black] (\a,0) circle (0.15);
        \foreach \a in {0.5,1.0,...,4.0}
            \fill[black] (\a,4.5) circle (0.15);
        \draw[thick,rounded corners=4] (3.5,0) -- (3.5,0.3) -- (4.0,0.3) -- (4.0,0);
        \draw[thick,rounded corners=4] (0.5,4.5) -- (0.5,4.2) -- (1.0,4.2) -- (1.0,4.5);
        \newcommand{\x}{3.0}
        \newcommand{\y}{0.9}
        \draw[thick,rounded corners=4] (\x,0) -- (\x,\y) -- (\x+0.5,\y) -- (\x+0.5,\y-0.3) -- (\x+1,\y-0.3) -- (\x+1,4.5);
        \renewcommand{\x}{2.5}
        \renewcommand{\y}{1.5}
        \draw[thick,rounded corners=4] (\x,0) -- (\x,\y) -- (\x+0.5,\y) -- (\x+0.5,\y-0.3) -- (\x+1,\y-0.3) -- (\x+1,4.5);
        \renewcommand{\x}{2.0}
        \renewcommand{\y}{2.1}
        \draw[thick,rounded corners=4] (\x,0) -- (\x,\y) -- (\x+0.5,\y) -- (\x+0.5,\y-0.3) -- (\x+1,\y-0.3) -- (\x+1,4.5);
        \renewcommand{\x}{1.5}
        \renewcommand{\y}{2.7}
        \draw[thick,rounded corners=4] (\x,0) -- (\x,\y) -- (\x+0.5,\y) -- (\x+0.5,\y-0.3) -- (\x+1,\y-0.3) -- (\x+1,4.5);
        \renewcommand{\x}{1.0}
        \renewcommand{\y}{3.3}
        \draw[thick,rounded corners=4] (\x,0) -- (\x,\y) -- (\x+0.5,\y) -- (\x+0.5,\y-0.3) -- (\x+1,\y-0.3) -- (\x+1,4.5);
        \renewcommand{\x}{0.5}
        \renewcommand{\y}{3.9}
        \draw[thick,rounded corners=4] (\x,0) -- (\x,\y) -- (\x+0.5,\y) -- (\x+0.5,\y-0.3) -- (\x+1,\y-0.3) -- (\x+1,4.5);
        \draw[dashed,rounded corners=4,line width=2] (4.25,0) -- (4.25,0.45) -- (3.25,0.45) -- (3.25,0.75) -- (3.75,0.75) -- (3.75,1.05) -- (2.75,1.05) -- (2.75,1.35) -- (3.25,1.35) -- (3.25,1.65) -- (2.25,1.65) -- (2.25,1.95) -- (2.75,1.95) -- (2.75,2.25) -- (1.75,2.25) -- (1.75,2.55) -- (2.25,2.55) -- (2.25,2.85) -- (1.25,2.85) -- (1.25,3.15) -- (1.75,3.15) -- (1.75,3.45) -- (0.75,3.45) -- (0.75,3.75) -- (1.25,3.75) -- (1.25,4.05) -- (0.25,4.05) -- (0.25,4.5);
        \draw[blue] (-0.3,4.5) node {$S_A$};
        \draw[blue] (4.8,4.5) node {$S_B$};
        \draw[blue] (-0.3,0.45) node {$6$};
        \draw[blue] (4.8,0.45) node {$0$};
        \draw[blue] (-0.3,1.05) node {$5$};
        \draw[blue] (4.8,1.05) node {$1$};
        \draw[blue] (-0.3,1.65) node {$4$};
        \draw[blue] (4.8,1.65) node {$2$};
        \draw[blue] (-0.3,2.25) node {$3$};
        \draw[blue] (4.8,2.25) node {$1$};
        \draw[blue] (-0.3,2.85) node {$2$};
        \draw[blue] (4.8,2.85) node {$0$};
        \draw[blue] (-0.3,3.45) node {$1$};
        \draw[blue] (4.8,3.45) node {$1$};
        \draw[blue] (-0.3,4.05) node {$0$};
        \draw[blue] (4.8,4.05) node {$0$};
        \end{tikzpicture}
    \end{minipage}
    }
    \caption{(Left) Teleportation protocol in the causal diagram of an evaporating black hole. Solid black lines are entangled Hawking quanta. Curved lines represent time slices counting the radiation entropy in region $B$ (number of lines crossing the slice). Solid blue lines depict teleported information -- they do not contribute to the entropy counting because they do not terminate in region $A$. (Right) Partial S matrix element (black) of the teleportation of a quantum state (blue). Dashed line is a separation of subsystems $A$ and $B$ drawn in such a way that particles can only enter region $A$, but not leave it. $S_A$ approximately computes the log of the number of microstates in $A$. $S_B$ approximately computes the entanglement entropy of the quanta in $A$ with those in $B$. The unshaded region of the interior is the initial island formed as the first piece of information is teleported at step four.}
    \label{fig:causald}
\end{figure}

The quantum teleportation protocol of type~(\ref{projection1}) can be embedded in a typical causal diagram of an evaporating black hole (figure~\ref{fig:causald}). Hawking quanta provide entangled pairs that can be used as a channel to extract information from the interior of the black hole by a distant observer. The necessary measurements of the protocol can be realized by particle interactions, as we are going to demonstrate.

An observer can measure the entropy of the Hawking radiation and construct an analog of the Page curve. The (fine-grained) entropy of radiation is roughly counted by the number of lines that cross a time slice outside of the horizon. If the teleportation protocol for a state in the interior is implemented, the lines representing the teleported state can be moved past the horizon and past the slice so that they no longer contribute to the count of the entropy~\cite{Melnikov:2022qyt}, as illustrated by the left panel of figure~\ref{fig:causald}. The topology of the connectivity changes, which is analogous to formation of an \emph{island} in the black hole interior~\cite{Almheiri:2019hni}. Heuristically the island is the void in lieu of the teleported bits, or the piece of the interior which was in the interior before, but was moved to the exterior via the teleportation protocol. Depending on whether one computes the entropy of the black hole, or the entropy of the radiation, one chooses one of these perspectives. In the first case, when all the Hawking quanta are utilized and the qubits are teleported, the island is the whole interior, and the entropy vanishes.

The naive teleportation protocol~(\ref{projection1}) is not unitary. One can note that neither the Hawking quanta creation process is unitary in the naive model. An obvious way to restore unitarity of the protocol is to have Hawking quanta created by the quanta annihilating in the interior. It is easy to imagine such a process in a topological model, like the one specified below. In this model, such a process can cross the horizon since breaking of the lines allows us to draw the horizon in between the annihilated and resurrected pair. Moreover, this should be true beyond the topological model since virtual particles close to the horizon are allowed to cross it, while on-shell particles can only cross the horizon in one direction. An example of the corresponding information extraction process is illustrated in the right panel of figure~\ref{fig:causald}.

The figure shows the result of a sequence of annihilation-creation processes happening in discrete time steps applied on one instance of the initial state, which is selected to represent a maximally non-local (scrambled) information storage inside the black hole. Such processes can happen, for example, under the evolution generated by a unitary braiding operator
\be
\begin{array}{c}
\scalebox{0.7}{\begin{tikzpicture}
\draw[line width=2.0,rounded corners=4] (0,0) -- (0,0.55) -- (0.75,0.85) -- (0.75,1.4);
\draw[line width=2,rounded corners=4] (0,1.4) -- (0,0.85) -- (0.25,0.75);
\draw[line width=2,rounded corners=4] (0.5,0.65) -- (0.75,0.55) -- (0.75,0);
\end{tikzpicture}}
\end{array}
\ = \ A
\begin{array}{c}
\scalebox{0.7}{\begin{tikzpicture}
\draw[line width=2,rounded corners=4] (0,0) -- (0.,1.4);
\draw[line width=2,rounded corners=4] (0.75,1.4) -- (0.75,0);
\end{tikzpicture}}
\end{array} \ + \ A^{-1}
\begin{array}{c}
\scalebox{0.7}{\begin{tikzpicture}
\draw[line width=2,rounded corners=4] (0,0) -- (0,0.55) -- (0.75,0.55) -- (0.75,0);
\draw[line width=2,rounded corners=4] (0,1.4) -- (0,0.85) -- (0.75,0.85) -- (0.75,1.4);
\end{tikzpicture}}
\end{array}\,,
\label{evolop}
\ee
where the first element on the right hand side is the identity operator and $A$ is a constant.\footnote{This is the Kauffman's version of the skein relation for braiding~\cite{Kauffman:1987bra}.} For simplicity, one can assume that this evolution applies only to the bits close to the dashed line (the horizon).

One can compute the evolution of the entropy in this process for both bits inside ($A$, coarse-grained black hole entropy) and outside ($B$, fine-grained radiation entropy) the horizon. These entropies, $S_A$ and $S_B$, are shown in figure~\ref{fig:causald} (right) for a particular choice of the initial state. Regardless of this choice, it is evident that in any such process, the entropy of the interior state $S_A$ (number of microstates) will always decrease. Meanwhile, the entropy $S_B$ (entanglement entropy) might initially increase, but it will always be bounded by $S_A$ and eventually vanish. This behavior aligns with the expected behavior of a unitary Page curve.

In the original work~\cite{Hawking:1976ra}, Hawking argues that the radiation is always maximally entangled with the interior (\emph{the principle of ignorance}). However, as explained by~\cite{Hayden:2007cs} and as the present model simply illustrates, this is only possible during the early stages of black hole evolution. After half of the black hole has evaporated (Page time), additional Hawking quanta will destroy the entanglement due to interactions necessarily present in the interior.

One should keep in mind that the black diagram in figure~(\ref{fig:causald}) (right) is not the full S matrix element as long as the evolution is generated by an operator like~(\ref{evolop}). The intermediate states of the evolution are linear combinations of states with and without radiation, for example, 
\be
\label{4partexample}
|\Psi\rangle \ = \ \alpha \begin{array}{c}
\scalebox{0.7}{\begin{tikzpicture}
\draw[line width=2,rounded corners=4,gray] (0.75,1.4) -- (0.75,0.65) -- (1.125,0.65);
\draw[line width=2,rounded corners=4,gray] (1.125,0.65) -- (1.5,0.65) -- (1.5,1.4);
\draw[line width=2,rounded corners=4,gray] (0,1.4) -- (0,0) -- (1.125,0.0);
\draw[line width=2,rounded corners=4,gray] (1.125,0) -- (2.25,0.0) -- (2.25,1.4);
\fill[gray] (0,1.4) circle (0.15);
\fill[gray] (0.75,1.4) circle (0.15);
\fill[gray] (1.5,1.4) circle (0.15);
\fill[gray] (2.25,1.4) circle (0.15);
\draw[line width=2.5,dashed] (2.6,-0.2) -- (2.6,1.6);
\end{tikzpicture}}
\end{array} 
+ \beta\begin{array}{c}
\scalebox{0.7}{\begin{tikzpicture}
\draw[line width=2,rounded corners=4,gray] (0.75,1.4) -- (0.75,0.65) -- (1.125,0.65);
\draw[line width=2,rounded corners=4,gray] (1.125,0.65) -- (1.5,0.65) -- (1.5,1.4);
\draw[line width=2,rounded corners=4,gray] (0,1.4) -- (0,0) -- (1.875,0.0);
\draw[line width=2,rounded corners=4] (1.875,0) -- (2.25,0.0) -- (2.25,1.4);
\fill[gray] (0,1.4) circle (0.15);
\fill[gray] (0.75,1.4) circle (0.15);
\fill[gray] (1.5,1.4) circle (0.15);
\fill[black] (2.25,1.4) circle (0.15);
\draw[line width=2.5,dashed] (1.875,-0.2) -- (1.875,1.6);
\end{tikzpicture}}
\end{array}
+ \gamma\begin{array}{c}
\scalebox{0.7}{\begin{tikzpicture}
\draw[line width=2,rounded corners=4,gray] (0.75,1.4) -- (0.75,0.65) -- (1.125,0.65);
\draw[line width=2,rounded corners=4] (1.125,0.65) -- (1.5,0.65) -- (1.5,1.4);
\draw[line width=2,rounded corners=4,gray] (0,1.4) -- (0,0) -- (1.125,0.0);
\draw[line width=2,rounded corners=4] (1.125,0) -- (2.25,0.0) -- (2.25,1.4);
\fill[gray] (0,1.4) circle (0.15);
\fill[gray] (0.75,1.4) circle (0.15);
\fill[black] (1.5,1.4) circle (0.15);
\fill[black] (2.25,1.4) circle (0.15);
\draw[line width=2.5,dashed] (1.125,-0.2) -- (1.125,1.6);
\end{tikzpicture}}
\end{array} \,,
\ee
where shaded parts of the diagrams correspond to the interior. In other words, the quanta outside the horizon only exist with a certain probability, and even the position of the horizon is not well-defined. The latter will be defined upon the measurement of the radiation. Practically, this means that the entropies will be computed for linear combinations of diagrams like the one in figure~\ref{fig:causald} (right). This will not affect the qualitative behavior of the Page curve, and the property $S_A\leq S_B$.

In summary, the considered toy model implements a topological version of the quantum teleportation protocol as the fundamental mechanism for extracting information from a causally disconnected region, analogous to the interior of a black hole. The model employs a simple interaction for both the teleportation and the production of the required entangled pairs (Hawking radiation). This topological model provides a straightforward illustration of the restriction on the growth of entropy for information outside the black hole, following a unitary Page curve. Additionally, it offers insights into the nature of the islands in the original entropy formula used in semiclassical gravity. The islands represent parts of the interior, which are in fact voids, corresponding to the modes ``teleported'' to the exterior. The term topological, used to define the present model, mostly refers to the description of quantum correlations in terms of lines connecting different parts of the system and ignoring its geometrical aspects.

\paragraph{Acknowledgments.} The author is grateful to Andrei Mironov, Sergey Mironov, Alexei Morozov and Andrey Morozov, with whom the ideas of the topological realization of quantum teleportation were originally discussed~\cite{M5}. This work was supported in part by the Simons Foundation grant number 884966-AF and the Serrapilheira Institute grant number Serra R-2012-38185. The present, significantly revised, version of the paper was produced with the additional support of the grant of Simons Foundation 1023171-RC and the grants 308580/2022-2 and 404274/2023-4 of the Brazilian Council for Scientific and Technological Developent (CNPq).

\end{document}